\documentclass{aastex701}

\usepackage{amsmath,amssymb}
\usepackage{bm}
\usepackage[rightcaption]{sidecap}

\providecommand{\He}{\alpha}
\providecommand{\mHe}{m_{\rm \alpha}}
\providecommand{\nHe}{n_{\rm \alpha}}
\providecommand{\nHeb}{n_{\rm \alpha,b}}
\providecommand{\RHe}{\mathcal{R}_{\rm \alpha}}
\providecommand{\THe}{T_{\rm \alpha}}
\providecommand{\VHe}{V_{\rm \alpha}}

\providecommand{\p}[1]{{\color{magenta}{#1}}}

\providecommand{\aap}{    {\it Astron. Astrophys.}}

\providecommand{\apjl}{   {\it Astrophys. J. Lett.}}

\usepackage{hyperref}

\begin{document}

\shorttitle{Ambipolar electric fields in multispecies atmospheres}
\shortauthors{Barbieri, D\'emoulin, \& Verscharen}

\title{The Ambipolar electric field in multispecies plasma atmospheres: effects of $\alpha$ particles and stochastic heating}

\author{Luca Barbieri}
\affiliation{LIRA, Observatoire de Paris, Universit\'e PSL, Sorbonne Universit\'e, Universit\'e Paris Cit\'e, CY Cergy Paris Universit\'e, CNRS, 92190 Meudon, France}
\email[show]{luca.barbieri@obspm.fr}
\correspondingauthor{Luca Barbieri}

\author{Pascal D\'emoulin}
\affiliation{LIRA, Observatoire de Paris, Universit\'e PSL, Sorbonne Universit\'e, Universit\'e Paris Cit\'e, CY Cergy Paris Universit\'e, CNRS, 92190 Meudon, France}
\email{} 

\author{Daniel Verscharen}
\affiliation{Mullard Space Science Laboratory, University College London, Dorking RH5 6NT, United Kingdom}
\email{} 

\begin{abstract}
We investigate stationary states of a collisionless gravitationally stratified plasma atmosphere composed of electrons, protons, and $\alpha$ particles by extending Pannekoek--Rosseland theory to multispecies and multi-temperature plasmas. Starting from Liouville's theorem, we derive the self-consistent ambipolar electric field from kinetic equilibrium and charge neutrality. For a single-temperature atmosphere, we obtain analytical expressions for the ambipolar field, demonstrate its strength depends on the $\alpha$ particle abundance, and derive the ordering of total potential energies and density profiles. A first-order analytical solution for the electrostatic potential reproduces the numerical solution with high accuracy. The formalism is then generalized to multi-temperature plasmas produced by stochastic boundary heating, in which the stationary velocity distribution is represented by a superposition of Maxwellian populations. Although stochastic heating substantially modifies the density and temperature profiles, the relative stratification of electrons, protons, and $\alpha$ particles remains unchanged, with $\alpha$ particles being the most strongly stratified and protons the least. The resulting atmosphere develops gravitational filtering, whereby low-energy particles are preferentially removed with increasing altitude, producing non-exponential density profiles and monotonically increasing temperatures without additional heating. The ambipolar electric field consists of a gravitational contribution, corresponding to the generalized Pannekoek--Rosseland field, and a thermoelectric contribution produced by species-dependent temperature gradients generated through gravitational filtering. While the gravitational component remains dominant, the thermoelectric term explains the non-monotonic structure of the total electric field. These results provide the basis for investigating the combined effects of plasma composition and stochastic heating on ambipolar electric fields in gravitationally stratified astrophysical plasmas.
\end{abstract}

\section{Introduction}

The atmospheres of the Sun and low-mass main-sequence stars,
together with stellar winds, are important
examples of astrophysical plasmas that depart from local thermodynamic
equilibrium \citep{Dudik_2017,Maksimovic_al_2020}. These departures occur in the form of 
significant deviations from Maxwellian velocity distributions, which are seen
both in spectroscopic observations of the solar atmosphere
\citep[e.g. ][]{Dudik_2017} and in \textit{in situ} measurements in the
solar wind \citep[e.g. ][]{Maksimovic_al_2020}. Over the past seven decades,
numerous kinetic models have therefore been developed to reproduce the
kinetic properties of the solar atmosphere and wind\citep{Chamberlain1960,Brandt1966,Jockers1970,Lemaire1971,Scudder1992a,Scudder1992b,Maksimovic1997,Landi-Pantellini2001,Vocks2001,Vocks_2002a,Vocks_2002b,Marsch2003,Vocks2003,Landi2003,Zouganelis2004,Vocks_2005,Marsch2006,Pierrard2011,Verscharen2019,Vocks2021,Jeong_2022,Barbieri2023temperature,barbieri2024temperaturedensityprofilescorona,Hau_2025,PtersdeBonhome2025,Barbieri2025c,Banik2026,Vinogradov2026}.

The solar atmosphere provides a natural environment in which to study
the effects of plasma composition on gravitational stratification and
ambipolar electric fields. The solar corona is composed predominantly
of electrons and protons, but also contains a significant population
of heavier ions, in particular fully ionized helium or $\alpha$ particles
\citep{Feldman2003}. The transition
from the partially ionized chromosphere to the fully ionized corona
occurs near the bottom of
the narrow transition region, only a few hundred
kilometres thick, where the temperature rises from approximately
\(10^4\,\mathrm{K}\) to \(10^6\,\mathrm{K}\)
\citep{GolubPasachoff:book,observedtemperature}.
In the transition region and the corona, the different plasma species are therefore subject simultaneously to the gravitational stratification and the ambipolar electric
field required to maintain charge-neutrality. Explaining the origin of
the rapid temperature increase remains one of the major unsolved
problems in solar physics and is commonly referred to as the coronal
heating problem
\citep{Klimchuk_2006,Reale2010,Klimchuk2015,VanDoorsselaere2020}.

Recent kinetic models based on gravitational filtering suggest that
stochastic heating  at the top of the chromosphere 
produces multi-temperature particle distributions that explain the observed density and temperature structure of the
upper solar atmosphere
\citep{Barbieri2023temperature,Barbieri2024b,
barbieri2024temperaturedensityprofilescorona,Barbieri2025c}. In these models, intermittent
heating events inject particle populations with different thermal
energies, while gravity preferentially removes the lower-energy
particles with increasing altitude. The resulting gravitational
filtering
\citep{Scudder1992a,Scudder1992b,Meyer-Vernet_2007} produces non-Maxwellian velocity distributions with a temperature profile that increases with height, even in the absence of
additional heating.

Previous studies generally adopt a two-component
electron--proton plasma. While this approximation captures the
fundamental mechanism responsible for the ambipolar electric field, it
neglects the contribution of heavier ion species. In the solar corona,
$\alpha$ particles typically accounts for several percent of the ion
population \citep{Feldman2003} and, owing to their larger
mass-to-charge ratio, they contribute to the gravitational
polarization of the plasma. The presence of $\alpha$ particles is
therefore expected to modify the ambipolar electric field, then the total
potential energies experienced by the different plasma species, and
consequently their density and temperature stratification.

Our goal is to extend the kinetic formalism of stochastic heating
developed in our previous work \citep{Barbieri2024b} to a multispecies
plasma composed of electrons, protons, and $\alpha$ particles. We investigate how plasma composition modifies the classical
Pannekoek--Rosseland equilibrium and how stochastic heating affects the
resulting ambipolar electric field, density profiles, and temperatures.

After introducing the collisionless multispecies kinetic model, 
we derive analytical expressions for the self-consistent ambipolar electric field in a
single-temperature atmosphere and investigate its dependence on the
$\alpha$-particle abundance (Section~\ref{sec:single-temperature}). We then
extend the formalism to multitemperature plasmas generated by stochastic
boundary heating, derive the corresponding analytical solutions, and
investigate the combined effects of plasma composition and
gravitational filtering on the density, temperature, and electrostatic
structure of the atmosphere (Section~\ref{sec:multi-temperature}).
Within the multitemperature framework, we also introduce a multifluid
formulation that decomposes the ambipolar electric field into
gravitational and thermoelectric contributions, providing a clear
physical interpretation of the kinetic results.

\section{Single-Temperature Solution}
\label{sec:single-temperature}

\subsection{Model Description}
\label{sec:model_description}

We consider a collisionless multispecies plasma atmosphere composed of electrons, protons, and $\alpha$ particles, assumed to be in a stationary configuration. Each particle species is subjected to the gravitational force
\begin{equation}
\mathbf{F}_{g,s}
=
m_{s}\mathbf{g},
\qquad
s\in\{e,p,\He\},
\end{equation}
where \(g=GM_{\odot}/R_{\odot}^{2}\) is the gravitational acceleration at the solar surface, \(G\) is the gravitational constant, and \(M_{\odot}\) and \(R_{\odot}\) denote the solar mass and radius, respectively. The quantity $m_s$ denotes the mass of a particle of species $s$. In addition to gravity, the plasma is immersed in an electrostatic field \(\mathbf{E}(z)\) and in a uniform magnetic field \(\mathbf{B}\) directed along the $z$-axis. Under these assumptions, the total force acting on particles of species \(s\) is
\begin{equation}
\label{totalforce}
\mathbf{F}_{s}
=
m_{s}\mathbf{g}
+
e_{s}
\left(
\mathbf{E}(z)
+
\frac{\mathbf{v}}{c}\times\mathbf{B}
\right),
\end{equation}
where \(c\) denotes the speed of light and $e_s$ is the electric charge of a particle of species $s$. Since the plasma is assumed to be collisionless, the distribution function is conserved along particle trajectories according to Liouville's theorem. The particle dynamics is therefore completely determined by the single-particle Hamiltonian
\begin{equation}
\label{hamiltonian}
\mathcal{H}_{s}
=
\frac{1}{2}m_{s}v^{2}
+
V_{s}(z),
\end{equation}
where the total potential is given by
\begin{equation}
\label{def_V}
V_{s}(z)
=
m_{s}\,g\,z
+
e_{s}\,\phi(z),
\end{equation}
and \(\phi(z)\) is the electrostatic potential, related to the electric field through \(\mathbf{E}=-\nabla\phi\) and satisfying the boundary condition \(\phi(0)=0\).

\subsection{Isothermal Kinetic Equilibrium}
\label{subsec:isothermal-equilibrium}

At the base of the plasma atmosphere (\(z=0\)), each plasma species is
assumed to be in thermal equilibrium with a heat reservoir characterized
by a common temperature \(T_b\). The corresponding boundary densities are
allowed to differ among the various species and are denoted by
\(n_{b,s}\). These densities are not independent but must satisfy
the charge-neutrality condition
\begin{equation}
\sum_{s\in\{e,p,\He\}}
e_s n_{b,s}=0.
\end{equation}
For an electron--proton--$\alpha$-particle plasma, this condition reduces to
\begin{equation}
\label{eq:boundary_neutrality}
n_{b,e}
=
n_{b,p}
+
2\,n_{b,\He},
\end{equation}
which guarantees charge neutrality at the lower boundary. Under these conditions, the equilibrium assumption specifies the
distribution function at the lower boundary, while Liouville's theorem
determines how this distribution is mapped throughout phase space along
particle trajectories. The resulting stationary distribution function
is therefore given by
\begin{equation}
\label{def_f}
\begin{aligned}
f_{T,s}(z,\mathbf v)
=
n_{b,s}
\left(
\frac{m_s}{2\pi k_B T_b}
\right)^{3/2}
\exp\!\left(
-\frac{\mathcal{H}_{s}}{k_B T_b}
\right),
\end{aligned}
\end{equation}
which corresponds to the Maxwell--Boltzmann equilibrium distribution
expressed in terms of the conserved single-particle Hamiltonian
\(\mathcal{H}_s\). Integrating Equation~\eqref{def_f} over velocity space yields the density profile of each species:
\begin{equation}
\label{densitysingletemp}
n_{s}(z)
=
n_{b,s}
\exp\!\left(
-\frac{V_{s}(z)}{k_B T_b}
\right).
\end{equation}
Since all
particles remain in thermal equilibrium with the reservoir, the kinetic temperature is spatially uniform,
\begin{equation}
\label{totaltemperaturepanne}
T(z)=T_b.
\end{equation}

\subsection{Ambipolar Field and the Pannekoek--Rosseland Limit}
\label{subsec:PR}

The electrostatic potential \(\phi(z)\) is determined by imposing charge-neutrality throughout the plasma,
\begin{equation}\label{eq:neutrality}
\sum_{s\in\{e,p,\He\}}
e_s n_s(z)
=
0.
\end{equation}

Differentiating the above condition with respect to \(z\) and using
Equation~\eqref{densitysingletemp}, we obtain a first-order differential equation for the electrostatic potential:
\begin{equation}
\label{eq:phi}
\frac{d\phi}{dz}
=
-g\,
\frac{
\displaystyle
\sum_s
e_s\, m_s\, n_s(z)
}{
\displaystyle
\sum_s
e_s^2\, n_s(z)
}.
\end{equation}
Equation~\eqref{eq:phi} provides a general expression for the
self-consistent ambipolar electric field in a multicomponent plasma. The
numerator represents the total gravitational force per unit volume,
weighted by the particle charge, whereas the denominator corresponds to
the collective electrostatic response of the plasma. The resulting
electric field therefore adjusts itself so as to maintain
charge-neutrality while balancing the different gravitational
stratifications of the charged species.

We note that, the same equilibrium electric field can be recovered within the multi-fluid formalism \citep{LEMAIRE1970103} for an open spherical geometry describing a polar wind. In that framework, additional terms arise in the electric field equation owing to the expanding geometry and plasma outflow. Here, instead, we derive the equilibrium electric field directly from the kinetic equilibrium distribution and the local charge-neutrality condition, without resorting to the fluid momentum equations. Furthermore, we investigate the Cartesian geometry in detail, highlighting its physical properties and deriving analytical solutions throughout the remainder of this section.

For a plasma composed of electrons, protons, and $\alpha$ particles,
Equation~\eqref{eq:phi} becomes
\begin{equation}
\label{eq:Fthree}
\frac{d\phi}{dz}
=
-\frac{g}{e}
\frac{
-m_en_e
+
m_pn_p
+
2\,\mHe \nHe
}{
n_e+n_p+4\,\nHe
}.
\end{equation}

As a consistency check, in the electron--proton limit
\(\nHe=0\), Equation~\eqref{eq:Fthree} can be integrated directly
using charge-neutrality. Imposing the boundary condition
\(\phi(0)=0\) yields
\begin{equation}
\phi(z)
=
-\frac{m_p-m_e}{2\,e}\,g\,z,
\label{eq:PRpotential}
\end{equation}
which is the classical Pannekoek--Rosseland electrostatic
potential derived for a gravitationally stratified two-species plasma
\citep{Pannekoek_1922,Rosseland_1924,Neslusan2001-rp,Barbieri_2025c}.

\subsection{Effect of $\alpha$ particles}
\label{subsec:alpha_effect}

The contribution of $\alpha$ particles can be made explicit by exploiting
the charge-neutrality condition given by Equation\eqref{eq:neutrality} which allows the electron density to be eliminated from
Equation~\eqref{eq:Fthree}:
\begin{equation}
\label{eq:Fthree_qn}
\frac{d\phi}{dz}
=
-\frac{g}{e}
\frac{
(m_p-m_e)\,n_p
+
2\,(\mHe -m_e)\,\nHe
}{
2\,n_p+6\,\nHe
}.
\end{equation}
Since the electron mass is negligible compared with the ion masses and
\(\mHe \simeq4\,m_p\), this expression simplifies to
\begin{equation}
\label{eq:Fapprox}
\frac{d\phi}{dz}
\simeq
-\frac{m_p\,g}{2\,e}
\frac{
n_p+8\,\nHe
}{
n_p+3\,\nHe
}.
\end{equation}

With the definition of the local
$\alpha$-particle abundance
\begin{equation}\label{Ionisationfactor}
f(z)
=
\frac{\nHe(z)}{n_p(z)},
\end{equation}
Equation~\eqref{eq:Fapprox} becomes
\begin{equation}
\label{eq:Ff}
\frac{d\phi}{dz}
\simeq
-\frac{m_p\,g}{2\,e}\,
\frac{1+8f}{1+3f}.
\end{equation}

The entire dependence of the ambipolar electric field on the plasma
composition is therefore contained in the dimensionless function
\begin{equation}
\mathcal{G}(f)
=
\frac{1+8f}{1+3f},
\end{equation}
which increases monotonically with the abundance of $\alpha$ particles. Moreover,
\begin{equation}
\lim_{f\rightarrow0}\mathcal{G}(f)=1,
\qquad
\lim_{f\rightarrow\infty}\mathcal{G}(f)=\frac{8}{3},
\end{equation}
so that
\begin{equation}
1
\le
\mathcal{G}(f)
<
\frac{8}{3}.
\end{equation}

Substituting these limits into Equation~\eqref{eq:Ff} immediately yields the
corresponding bounds for the ambipolar electric field,
\begin{equation}\label{boundssingletemp}
\frac{m_p\,g}{2\,e}
\le
-\frac{d\phi}{dz}
<
\frac{4\,m_p\,g}{3\,e},
\end{equation}
the ambipolar electric field therefore remains bounded between the electron--proton and $\alpha$-particle-dominated limits, with the plasma composition determining the exact value of the electrostatic potential gradient. Then, the classical Pannekoek--Rosseland solution represents the minimum ambipolar field, $-d\phi/dz$, compatible with charge-neutrality in a gravitationally stratified plasma. The bounds derived above also determine the ordering of the total potential energies of the different plasma species. Neglecting the electron mass, the total potential gradients are
\begin{equation}
\label{derivativepotentialenergies}
\begin{aligned}
\frac{dV_e}{dz}
&=
-e\frac{d\phi}{dz},
\\
\frac{dV_p}{dz}
&=
m_p g
+
e\frac{d\phi}{dz},
\\
\frac{d\VHe}{dz}
&=
4\,m_p g
+
2\,e\frac{d\phi}{dz}.
\end{aligned}
\end{equation}

Using the bounds on the ambipolar electric field given by
Equation~\eqref{boundssingletemp}, together with the classical
Pannekoek--Rosseland result
\begin{equation}
\frac{dV_p^{\rm PR}}{dz}
=
\frac{dV_e^{\rm PR}}{dz}
=
\frac{m_p\,g}{2},
\end{equation}
we obtain
\begin{equation}
\frac{dV_p}{dz}
<
\frac{dV_p^{\rm PR}}{dz}
=
\frac{dV_e^{\rm PR}}{dz}
<
\frac{dV_e}{dz}
<
\frac{d\VHe}{dz}.
\end{equation}

Since all total potentials vanish at the lower boundary, the same
ordering is preserved throughout the atmosphere:
\begin{equation}\label{eq:potential_ordering}
V_p
<
V_p^{\rm PR}
=
V_e^{\rm PR}
<
V_e
<
\VHe.
\end{equation}
The ordering of the total potential energies immediately determines the relative density stratification. Since the densities satisfy Equation \eqref{densitysingletemp}, the normalized densities satisfy
\begin{equation}\label{eq:ordering_densities}
\frac{n_p}{n_{p,b}}
>
\frac{n_p^{\rm PR}}{n_{p,b}}
=
\frac{n_e^{\rm PR}}{n_{e,b}}
>
\frac{n_e}{n_{e,b}}
>
\frac{\nHe}{\nHeb}.
\end{equation}

Therefore, the enhanced ambipolar electric field generated by $\alpha$ particles reduces the gravitational stratification of protons while increasing that of electrons. $\alpha$ particles remain the most strongly stratified species because of their large total potential energy.

\subsection{Analytical Solution for the Ambipolar Potential}
\label{subsec:local_analytical_solution}

Equation~\eqref{eq:phi} is, in general, a nonlinear ordinary
differential equation subject to the boundary condition
\(\phi(0)=0\) and does not admit a closed-form analytical
solution. Nevertheless, an approximate analytical solution can be
obtained sufficiently close to the lower boundary where
\begin{equation}
\left|
\frac{V_s(z)}{k_BT_b}
\right|
\ll 1
\label{eq:small_potential_condition}
\end{equation}
for all plasma species. This condition states that the
gravitational and electrostatic potential energies remain much smaller
than the thermal energy of every species. Consequently, thermal motion
dominates over the effects of stratification, allowing the Boltzmann
factors to be expanded about the lower boundary as
\begin{equation}
\exp\!\left(
-\frac{V_s}{k_BT_b}
\right)
\simeq
1-
\frac{V_s}{k_BT_b}.
\label{eq:local_exponential_expansion}
\end{equation}

\subsubsection{Zeroth-Order Solution}

At leading order, the exponential factors entering
Equation~\eqref{eq:phi} can be approximated by unity. The ambipolar
electric field therefore becomes constant, and the corresponding
electrostatic potential is
\begin{equation}
\phi(z)
\simeq
F_0\, z,
\label{eq:local_phi_general}
\end{equation}
where
\begin{equation}
F_0
=
-g
\frac{
\displaystyle
\sum_s e_s\, m_s\, n_{b,s}
}{
\displaystyle
\sum_s e_s^2\, n_{b,s}
},
\label{eq:local_field_general}
\end{equation}
is the value of the ambipolar field evaluated at the lower boundary.
With the introduction of the effective mass
\begin{equation}
\mathcal{M}(f_b)
=
\frac{1}{2}
\frac{
(m_p-m_e)
+
2\,f_b(\mHe -m_e)
}{
1+3\,f_b
},
\label{eq:effective_ambipolar_mass}
\end{equation}
where \(f_b = f(0) \) is the $\alpha$-particle abundance at the lower boundary,
defined by Equation~\eqref{Ionisationfactor}, the electrostatic potential writes
\begin{equation}
\phi(z)
\simeq
-\frac{g\,z}{e}\mathcal{M}(f_b).
\label{eq:local_phi_effective_mass}
\end{equation}
In the absence of $\alpha$ particles (\(f_b=0\)), this expression reduces
to the classical Pannekoek--Rosseland potential given by
Equation~\eqref{eq:PRpotential}. The total potential energy experienced by each plasma species follows directly
from Equation~\eqref{def_V}. Combining Equations~\eqref{eq:local_phi_effective_mass}
and~\eqref{def_V} yields
\begin{equation}
V_e(z)
=
\left[
m_e+\mathcal{M}(f_b)
\right]g\,z,
\label{eq:electron_total_potential}
\end{equation}
\begin{equation}
V_p(z)
=
\left[
m_p-\mathcal{M}(f_b)
\right]g\,z,
\label{eq:proton_total_potential}
\end{equation}
\begin{equation}
\VHe (z)
=
\left[
\mHe -2\mathcal{M}(f_b)
\right]g\,z.
\label{eq:helium_total_potential}
\end{equation}
These expressions show that the ambipolar electric field redistributes
the effective confinement of the different plasma species according to
both their masses and their electric charges.


\subsubsection{First-Order Solution}

Retaining the first-order terms in the expansion of the exponential factors within Equation~\eqref{eq:phi} accounts for the gradual variation of the species densities and therefore introduces the leading spatial correction to the ambipolar electric field. As a consequence, both the electric field and the electrostatic potential become nonlinear functions of height. Expanding the exponential factors to first order as given by Equation \eqref{eq:local_exponential_expansion} and substituting into Equation~\eqref{eq:phi}, yields the
linear differential equation
\begin{equation}
\frac{dy}{dz}
-
\frac{c_b}{H_b}y
=
-a_b
+
\frac{b_b}{H_b}z,
\label{eq:first_order_ode}
\end{equation}
where
\begin{equation}
y=\frac{e\,\phi}{m_p\,g},
\end{equation}
\begin{equation}
H_b=\frac{k_BT_b}{m_p\,g},
\end{equation}
\begin{equation}
a_b
=
\frac{1+8\,f_b}
{2\,(1+3\,f_b)},
\end{equation}
\begin{equation}
b_b
=
\frac{1+46\,f_b+64\,f_b^2}
{4\,(1+3\,f_b)^2},
\end{equation}
\begin{equation}
c_b
=
\frac{1+16\,f_b+24\,f_b^2}
{2\,(1+3\,f_b)^2}.
\end{equation}

The corresponding analytical solution is
\begin{equation}
\phi(z)
=
\frac{m_p\,g\,H_b}{e}
\Bigg[
-\frac{a_b}{c_b}
\left(
e^{c_bz/H_b}-1
\right)
+\frac{b_b}{c_b^2}
\left(
e^{c_bz/H_b}
-
1
-
\frac{c_bz}{H_b}
\right)
\Bigg],
\label{eq:first_order_phi}
\end{equation}
which reduces to the leading-order expression
(Equation~\eqref{eq:local_phi_effective_mass}) in the limit
\(z/H_b\ll1\). The total potential energies follow directly from
Equation~\eqref{def_V}, and therefore inherit the nonlinear dependence of the
electrostatic potential on height. Consequently, the density profiles are no
longer described by simple exponential functions with constant scale
heights, but instead reflect the progressive variation of the ambipolar
electric field. 

\begin{figure*}
\centering
\includegraphics[width=0.99\textwidth]{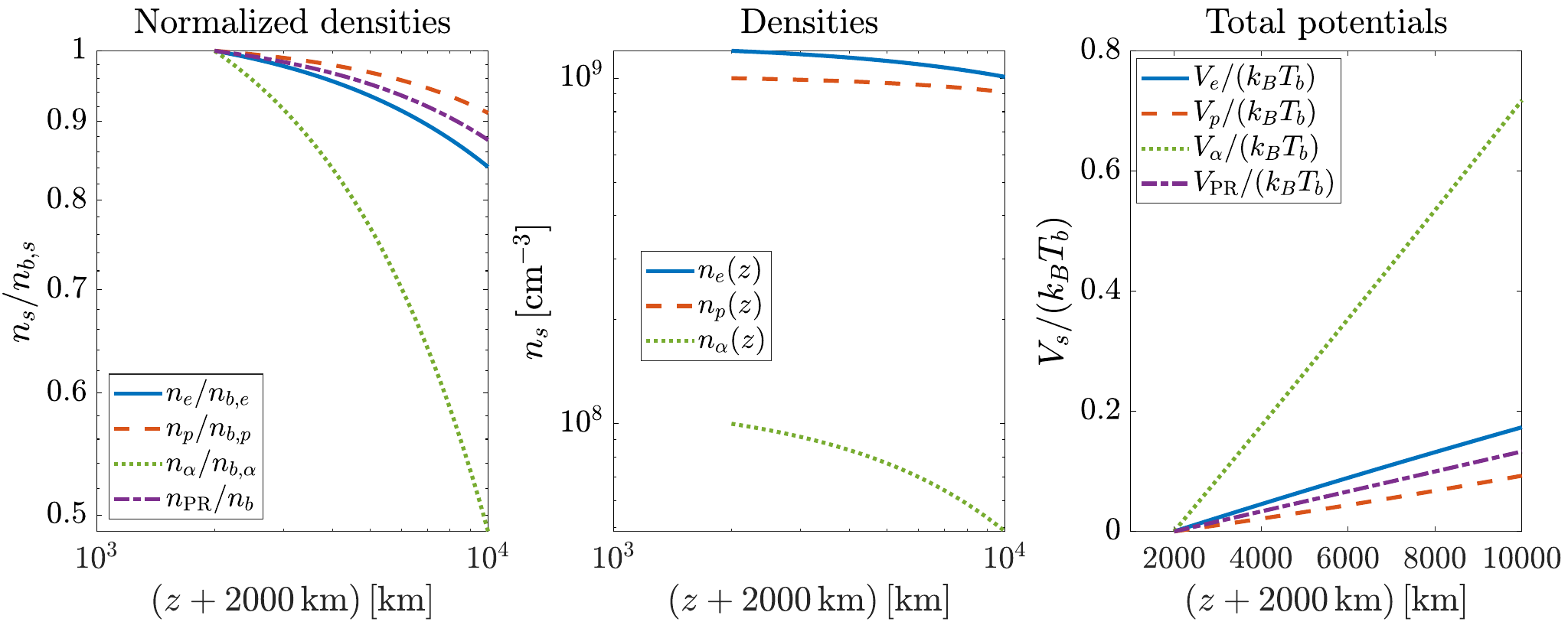}
\caption{
Density and total potential profiles for a three-component plasma
composed of electrons, protons, and $\alpha$-particles.
\textit{Left panel:} normalized density profiles of electrons,
protons, and $\alpha$ particles together with the classical
Pannekoek--Rosseland solution as functions of height. The density
profiles are computed from Equation~\eqref{densitysingletemp} using the
self-consistent electrostatic potential obtained by numerically solving
Equation~\eqref{eq:Fthree}, whereas the Pannekoek--Rosseland reference
profile is computed from
Equation~\eqref{eq:PRpotential}.
\textit{Middle panel:} corresponding density profiles.
\textit{Right panel:} total potential energies, computed from
Equation~\eqref{def_V} and normalized to \(k_BT_b\), as functions of height.
The calculations are performed assuming 
\(T_b=10^6\,\mathrm{K}\), 
\(f_b=n_{b,\He}/n_{b,p}=0.1\), and 
\(
n_{b,p}=10^9\,\mathrm{cm^{-3}}\).
}
\label{fig1}
\end{figure*}

\begin{figure}
\centering
\includegraphics[width=0.45\columnwidth ]{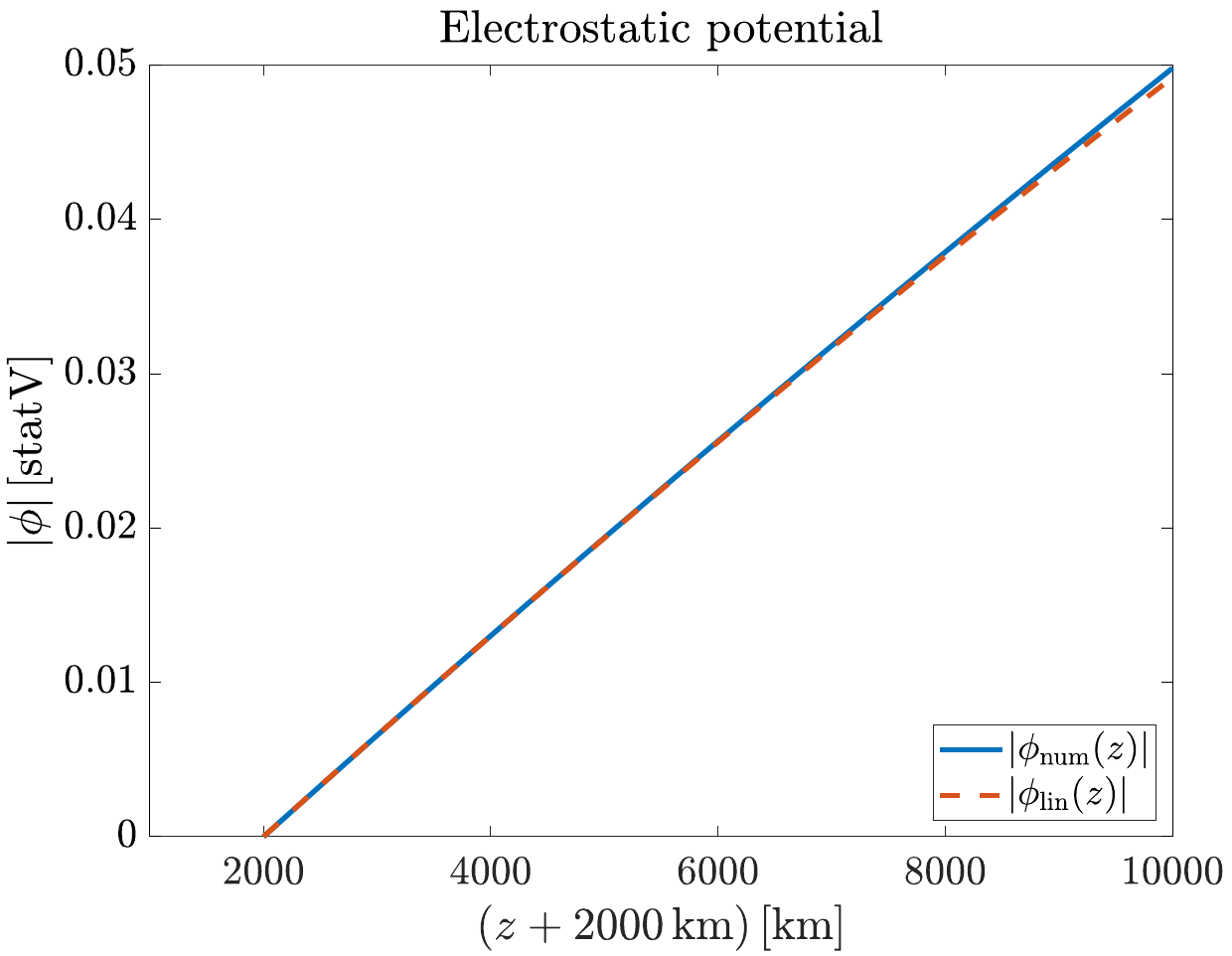}
\caption{
Comparison between the numerical solution of the nonlinear
self-consistent electrostatic potential obtained from
Equation~\eqref{eq:Fthree} and the first-order analytical approximation
given by Equation~\eqref{eq:first_order_phi}. The calculations were
performed using the same physical parameters as in
Figure~\ref{fig1}.
}
\label{fig2}
\end{figure}

\subsection{Density, Total, and Electrostatic Potential Profiles}

In this section we present the density, total potential, and
electrostatic potential profiles for a single-temperature
three-component plasma. We obtain the electrostatic potential by
solving numerically Equation~\eqref{eq:Fthree} using a fourth-order Runge--Kutta scheme subject to the boundary condition
\(\phi(0)=0\).

Figure~\ref{fig1} summarizes the main results. The right panel shows the total potential energies experienced by the
different species. As predicted by the leading-order analytical
solution, the total potentials remain nearly linear functions of height and satisfy the ordering given by Equation~\eqref{eq:potential_ordering}. The ambipolar
electric field reduces the effective gravitational force acting on all charged species, albeit by different amounts. $\alpha$ particles  experiences the largest total potential owing to its larger mass-to-charge ratio, whereas protons experience the smallest one. The left panel displays the corresponding normalized density profiles.
Their ordering is opposite to the potential ordering and is
therefore consistent with Equation~\eqref{eq:ordering_densities}. Protons possess the largest effective scale height, whereas $\alpha$ particles exhibit the
strongest gravitational stratification. The electron density lies between the proton and $\alpha$ particles profiles because the ambipolar electric field partially compensates gravity while simultaneously preserving charge-neutrality. The central panel shows the corresponding density profiles.
The electron and proton densities remain close to each other, as a consequence of the charge-neutrality constraint and the relatively low $\alpha$-particle abundance. In contrast, the $\alpha$-particle density decreases more rapidly with height because of its smaller effective gravitational scale height. Figure~\ref{fig1} therefore illustrates the
direct connection between the species-dependent total potentials and their density stratification: species experiencing steeper total potentials exhibit more rapidly decreasing density profiles, as set by the self-consistent ambipolar electric field in the multicomponent plasma.

Figure~\ref{fig2} compares the first-order analytical solution for the electrostatic potential, Equation~\eqref{eq:first_order_phi}, with the corresponding numerical solution of the full nonlinear problem. The
agreement is excellent over the entire computational domain, with only very small deviations appearing near the upper limit. This behaviour is fully consistent with the assumptions underlying the analytical approximation. The first-order solution is obtained by expanding the exponential dependence of the density profiles with respect to the dimensionless parameter
\(V_s/(k_BT_b)\), as expressed by Equation~\eqref{eq:small_potential_condition}. As shown in the right panel of Figure~\ref{fig1}, the normalized total potentials remain smaller than unity for all plasma species throughout the computational domain, so that the neglected higher-order terms remain small. Therefore, the first-order approximation accurately reproduces the exact nonlinear solution over the entire height range considered. Nonlinear corrections become progressively more important with increasing altitude, where the normalized total potentials approach unity.

\section{Multi-Temperature Solution}
\label{sec:multi-temperature}

\subsection{Stochastic Boundary Heating}
\label{subsec:stochastic-boundary-heating}

We now extend the kinetic formalism developed in the previous sections  to boundary conditions characterized by stochastic temperature
fluctuations rather than a single Maxwellian distribution. Here we
consider the multi-temperature model introduced by
\citet{Barbieri2023temperature,Barbieri2024b}, in which the boundary is
located at the top of the chromosphere and undergoes intermittent
heating events. During a heating event, the boundary temperature increases from the
background chromospheric value \(T_0\simeq10^4\,\mathrm{K}\) to
\(T=T_0+\Delta T\), where the heated temperature is sampled from a
normalized probability density \(\gamma(T)\). Heating episodes occupy a
fraction
\begin{equation}
A_t=\frac{\tau}{\tau+t_w}
\end{equation}
of the total time, where \(\tau\) is the duration of each event and
\(t_w\) is the waiting time between consecutive events. As in
\citet{Barbieri2023temperature,Barbieri2024b}, we focus on
the regime
\begin{equation}\label{shorttimescalesregime}
\tau,\;t_w\ll t_{R,e},
\end{equation}
for which successive heating events occur faster than the plasma
relaxation time $t_{R,e}$. The boundary therefore continuously injects particle
populations with different temperatures, leading to a stationary
multi-temperature distribution whose statistical properties are
determined by \(A_t\) and \(\gamma(T)\).

\subsection{Stationary Distribution and Velocity Moments}
\label{subsec:multi-temperature-distribution}

For the short-timescale regime defined by
Equation~\eqref{shorttimescalesregime},
\citet{Barbieri2024b,Barbieri2025b} show that the stationary velocity
distribution function of species \(s\) is
\begin{equation}
\label{VDFphasespace}
f_{s,M}(z,\mathbf{v})
=
\mathcal{N}_{s}
\Bigg[
A_t
\int_{T_0}^{+\infty}
\frac{\gamma(T)}{T^{3/2}}
\exp\!\left[
-\frac{\mathcal{H}_{s}}{k_B T}
\right]
\,dT
+
\frac{1-A_t}{T_0^{3/2}}
\exp\!\left[
-\frac{\mathcal{H}_{s}}{k_B T_0}
\right]
\Bigg],
\end{equation}
where the normalization constant is
\begin{equation}
\label{eq:multiT_normalization}
\mathcal{N}_{s}
=
n_{0,s}
\left(
\frac{m_s}{2\pi k_B}
\right)^{3/2},
\end{equation}
and \(n_{0,s}\) is the boundary density of species \(s\).
The boundary densities must satisfy charge-neutrality given by Equation \eqref{eq:boundary_neutrality}. The distribution in Equation~\eqref{VDFphasespace} is a weighted
superposition of Maxwellian populations. The first term describes the
ensemble of heated states generated by the temperature distribution
\(\gamma(T)\), whereas the second term represents the background
population with temperature \(T_0\). The density profile follows from the zeroth-order velocity moment:
\begin{equation}
\label{density_multiT}
\begin{aligned}
n_{s,M}(z)
=
n_{0,s}
\Bigg[
A_t
\int_{T_0}^{+\infty}
\gamma(T)
\exp\!\left[
-\frac{V_s(z)}{k_B T}
\right]
\,dT
+
(1-A_t)
\exp\!\left[
-\frac{V_s(z)}{k_B T_0}
\right]
\Bigg].
\end{aligned}
\end{equation}

The kinetic temperature follows from the second-order velocity moment:
\begin{equation}
\label{paralleltemperaturepanne2}
\begin{aligned}
T_{s,M}(z)
=
\frac{n_{0,s}}{n_{s,M}(z)}
\Bigg[
A_t
\int_{T_0}^{+\infty}
\gamma(T)\,T
\exp\!\left[
-\frac{V_s(z)}{k_B T}
\right]
\,dT
+
(1-A_t)\,T_0
\exp\!\left[
-\frac{V_s(z)}{k_B T_0}
\right]
\Bigg].
\end{aligned}
\end{equation}

Equations~\eqref{density_multiT} and
\eqref{paralleltemperaturepanne2} reflect that the density and temperature
at a given height are determined by different moments of the stochastic
temperature distribution. The density weights each thermal population
through its Boltzmann factor, whereas the kinetic temperature contains an
additional factor \(T\).

\subsection{Self-Consistent Multi-Temperature Ambipolar Field}
\label{subsec:multi-temperature-ambipolar-field}

As in the single-temperature case,
Equations~\eqref{density_multiT} and
\eqref{paralleltemperaturepanne2} are not explicit because
\(V_s(z)\) depends on the self-consistent electrostatic potential
\(\phi(z)\). The latter is determined by imposing charge-neutrality given by Equation \eqref{eq:neutrality}.
As for the single-particle case, we impose charge neutrality via Equation \eqref{eq:neutrality}. Using the densities given by Equation \eqref{density_multiT} and differentiating the resulting equation, we find
\begin{equation}
\label{eq:phi_multiT}
\frac{d\phi}{dz}
= -g
\frac{
\displaystyle
\sum_s
e_s\, m_s\, n_{0,s}\,
\mathcal{R}_s(\phi,z)
}{
\displaystyle
\sum_s
e_s^2\, n_{0,s}\,
\mathcal{R}_s(\phi,z)
},
\end{equation}
where the multi-temperature response function $\mathcal{R}_s(\phi,z)$ is defined as
\begin{equation}
\label{eq:thermal_response}
\begin{split}
\mathcal{R}_s(\phi,z)
=
n_{0,s}
\Bigg(A_t
\int_{T_0}^{+\infty}
\frac{\gamma(T)}{T}
\exp\!\left[
-\frac{V_s(z)}{k_BT}
\right]
\,dT
+\frac{1-A_t}{T_0}
\exp\!\left[
-\frac{V_s(z)}{k_BT_0}
\right]
\Bigg).
\end{split}
\end{equation}

For a plasma composed of electrons, protons, and $\alpha$ particles,
Equation~\eqref{eq:phi_multiT} becomes
\begin{equation}
\label{eq:FM_three_species}
\frac{d\phi}{dz}
=
-\frac{g}{e}\,
\frac{
m_p\mathcal{R}_p
+
2\,\mHe \RHe
-m_e\mathcal{R}_e
}{
\mathcal{R}_e
+
\mathcal{R}_p
+
4\,\RHe
}.
\end{equation}
The single-temperature limit is recovered by setting
\(\gamma(T)=\delta(T-T_b)\) and \(A_t=1\), in which case
Equation~\eqref{eq:phi_multiT} reduces identically to Equation \eqref{eq:phi}. The main difference with respect to the single-temperature case is that
the contribution of each species is no longer determined directly by its
local density. Instead, it is governed by the response function
\(\mathcal{R}_s\), which contains the weighted superposition of all
the temperature populations generated by the stochastic heating process.
The self-consistent electrostatic field thus retains information about the
thermal history imposed at the lower boundary.

\subsubsection{Electron--Proton Limit}

In the electron--proton limit,
\(\RHe=0\),
Equation~\eqref{eq:FM_three_species} becomes
\begin{equation}
\label{eq:FM_electron_proton}
\frac{d\phi}{dz}
=
-\frac{g}{e}
\frac{
m_p\mathcal{R}_p
-
m_e\mathcal{R}_e
}{
\mathcal{R}_e+\mathcal{R}_p
}.
\end{equation}

This expression represents the multi-temperature generalization of the
Pannekoek--Rosseland electric field. As demonstrated by  \citet{barbieri2026b}, charge-neutrality in a multi-temperature
electron--proton plasma implies $\mathcal{R}_e = \mathcal{R}_p$.
Then, Equation~\eqref{eq:FM_electron_proton} recovers the classical
Pannekoek--Rosseland electrostatic potential given by
Equation~\eqref{eq:PRpotential}. 

\subsubsection{Effect of $\alpha$ particles}

We now investigate the effect of including $\alpha$ particles in the
multi-temperature model. Their contribution enters both the numerator
and the denominator of Equation~\eqref{eq:FM_three_species}, enhancing
the gravitational electric response while simultaneously increasing
the electrostatic response of the plasma. Neglecting the electron mass
and using \(m_{\rm He}\simeq4\,m_p\),
Equation~\eqref{eq:FM_three_species} becomes
\begin{equation}
-\frac{d\phi}{dz}
=
\frac{m_p\,g}{e}
\frac{
\mathcal{R}_p
+
8\,\mathcal{R}_{\rm \He}
}{
\mathcal{R}_e
+
\mathcal{R}_p
+
4\,\mathcal{R}_{\rm \He}
}.
\label{eq:FM_three_species_mass_approx}
\end{equation}
Since all response functions are positive,
\begin{equation}
\mathcal{R}_p+8\,\mathcal{R}_{\rm \He}
<
2\left(
\mathcal{R}_e+\mathcal{R}_p+4\,\mathcal{R}_{\rm \He}
\right),
\end{equation}
and therefore the ambipolar electric field always satisfies the upper
bound
\begin{equation}
-\frac{d\phi}{dz}
<
\frac{2\,m_p\,g}{e}.
\label{eq:upper_bound_multitemp}
\end{equation}

A lower bound can be obtained in the physically relevant regime of a
small $\alpha$-particle abundance. Under typical solar conditions, the
$\alpha$-particle population represents only a small correction to the
electron--proton plasma, so that the electron and proton responses
remain approximately equal,
\(\mathcal{R}_e\simeq\mathcal{R}_p\). This approximation is further
justified below through the perturbative expansion in the
$\alpha$-particle abundance. Equation~\eqref{eq:FM_three_species_mass_approx}
then reduces to
\begin{equation}
\label{eq:FM_three_species_reduced}
\frac{d\phi}{dz}
=
-\frac{m_p\,g}{2\,e}
\frac{
\mathcal{R}_p
+
8\,\mathcal{R}_{\rm \He}
}{
\mathcal{R}_p
+
2\,\mathcal{R}_{\rm \He}
}.
\end{equation}
Introducing
\(
\eta=\mathcal{R}_{\rm \He}/\mathcal{R}_p
\),
the composition-dependent factor becomes $(1+8\eta)\,/\,(1+2\eta)$
%
%
which increases monotonically from unity at \(\eta=0\) to \(4\) in
the alpha-dominated limit. Consequently,
\begin{equation}
\label{boundsmultitemp}
\frac{m_p\,g}{2\,e}
\leq
-\frac{d\phi}{dz}
<
\frac{2\,m_p\,g}{e}.
\end{equation}
Thus, the $\alpha$-particle-dominated limit provides a general upper bound on the strength of the ambipolar electric field, whereas, in the
small-alpha-abundance regime where \(\mathcal{R}_e\simeq\mathcal{R}_p\), the electron--proton limit
recovers the Pannekoek--Rosseland field as the lower bound. Unlike in the single-temperature case, however, the relevant control parameter is no longer solely the alpha-to-proton abundance ratio, but the response-function ratio
\(\mathcal{R}_{\rm \He}/\mathcal{R}_p\). This quantity depends on both the plasma composition and the parameters characterizing the stochastic heating process, and generally varies with height through the thermal response functions. Consequently, the enhancement of the ambipolar electric field is controlled jointly by the species abundances and the
statistical properties of the boundary heating.

From the bounds in Equation~\eqref{boundsmultitemp}, we derive the ordering of the total potential energies of the different plasma species by following the same procedure adopted for the single-temperature atmosphere. The resulting ordering is therefore identical to that obtained in the previous section and given by Equation~\eqref{eq:potential_ordering}.
Since the density of each species is a monotonically decreasing function of its total potential energy, the hierarchy given by Equation~\eqref{eq:potential_ordering} directly leads to the corresponding ordering of the normalized densities given by Equation \eqref{eq:ordering_densities}, independently of the adopted temperature distribution.

The main difference in the multi-temperature case concerns the
temperature profiles. Owing to gravitational filtering, the steeper the total potential of a given species, the stronger is the preferential selection of higher-energy particles at larger heights. The ordering of the total potential energies given by Equation~\eqref{eq:potential_ordering} then directly leads to the corresponding ordering of the effective temperatures as

\begin{equation}\label{eq:temperatures_ordering}
T_p
<
T_p^{\rm PR}
=
T_e^{\rm PR}
<
T_e
<
\THe \,.
\end{equation}

\subsection{Multi-Fluid Description of the Multi-Temperature Plasma} 
\label{sec:multi-fluid_electric_field}

We now interpret the effect of a multi-temperature plasma on the ambipolar electric field through a multi-fluid description. Since the
particle velocity distribution functions given by
Equation~\eqref{VDFphasespace} are symmetric in velocity space, the bulk
velocity of each species vanishes together with the interspecies
friction force. The momentum equation for a generic species therefore
reduces to
\begin{equation}
\frac{dp_s}{dz}
=
-m_s\, g\, n_s
+
e_s\, E\, n_s,
\label{eq:multi-fluid_momentum}
\end{equation}
where \(p_s\) denotes the pressure of species \(s\). Using
the kinetic definition of  pressure,
\begin{equation}
p_s
=
n_s k_B T_s,
\label{eq:multi-fluid_pressure}
\end{equation}
the momentum equation becomes
\begin{equation}
\frac{dn_s}{dz}
=
n_s
\left[
\frac{e_s E-m_s g}
{k_BT_s}
-
\frac{d\ln T_s}{dz}
\right].
\label{eq:multi-fluid_density_gradient}
\end{equation}

Differentiating the charge-neutrality condition in
Equation~\eqref{eq:neutrality} with respect to \(z\) and substituting
Equation~\eqref{eq:multi-fluid_density_gradient} yields
\begin{equation}
E(z)
=
E_{\rm g}(z)
+
E_{\rm T}(z),
\label{eq:E_multi-fluid_decomposition}
\end{equation}
where
\begin{equation}
E_{\rm g}(z)
=
g\,
\frac{
\displaystyle
\sum_s
e_s m_s n_s(z)
\,/ T_s(z)
}{
\displaystyle
\sum_s
e_s^2 \,n_s(z)
\,/ T_s(z)
}
\label{eq:E_multi-fluid_gravitational}
\end{equation}
is the gravitational contribution to the ambipolar electric field, while
\begin{equation}
E_{\rm T}(z)
=
k_B
\frac{
\displaystyle
\sum_s
e_s\, n_s(z)
\,d\ln T_s / 
dz
}{
\displaystyle
\sum_s
e_s^2\, n_s(z)
/ T_s(z)
}
\label{eq:E_multi-fluid_thermal}
\end{equation}
is the thermoelectric contribution associated with
species-dependent temperature gradients. If all species share the same temperature profile, $T_s(z)=T(z) \quad \forall\,s$,
the common temperature in Equation~\eqref{eq:E_multi-fluid_gravitational} cancels, which immediately reduces to
Equation~\eqref{eq:phi}. Moreover, the thermoelectric contribution
becomes
\begin{equation}
E_{\rm T}
=
k_B
\frac{
\dfrac{d T}{dz}
\displaystyle\sum_s e_s\, n_s
}{
\displaystyle
\sum_s
e_s^2\, n_s
}
=
0,
\end{equation}
where the last equality follows directly from the charge-neutrality
condition in Equation~\eqref{eq:neutrality}. Therefore, a non-vanishing
thermoelectric contribution requires different temperature gradients
for the different plasma species.

\begin{figure*}
\centering
\includegraphics[width=0.99\textwidth]{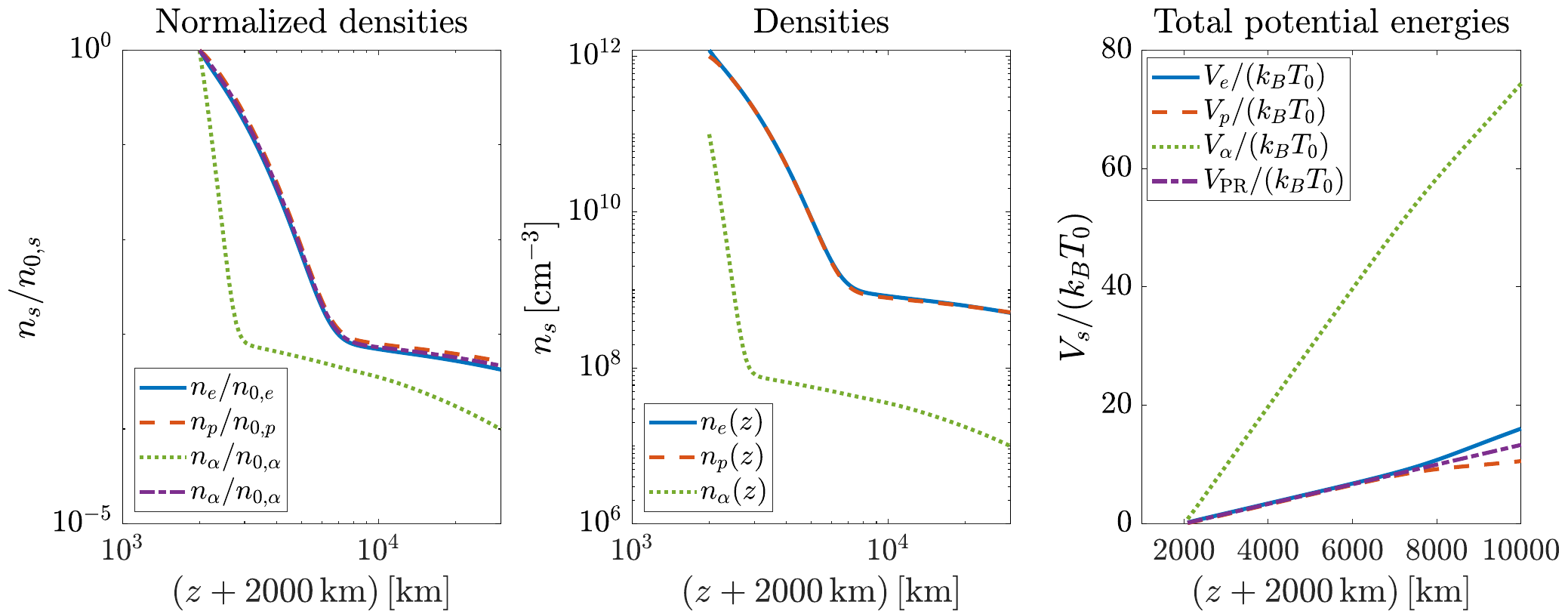}
\caption{
Density and total potential profiles for a multi-temperature
three-component plasma.
\textit{Left panel:} normalized density profiles of electrons,
protons, $\alpha$ particles, and the corresponding
Pannekoek--Rosseland solution as functions of height. The density
profiles are computed from Equation~\eqref{density_multiT} using the
self-consistent electrostatic potential obtained by numerically
solving Equation~\eqref{eq:FM_three_species}.
\textit{Middle panel:} corresponding density profiles.
\textit{Right panel:} total potential energies, computed from
Equation~\eqref{def_V} and normalized to \(k_BT_0\), as functions of
height. The calculations are performed
assuming \(T_0=10^4\,\mathrm{K}\), \(\Delta T = 10^{6} \mathrm{K}\), \(A_t = 10^{-3}\),
\(f_b=n_{0,\He}/n_{0,p}=0.1\), and 
\(
n_{0,p}=10^{12}\,\mathrm{cm^{-3}}\).
}
\label{fig3}
\end{figure*}

\begin{figure}
\centering
\includegraphics[width=0.5\columnwidth]{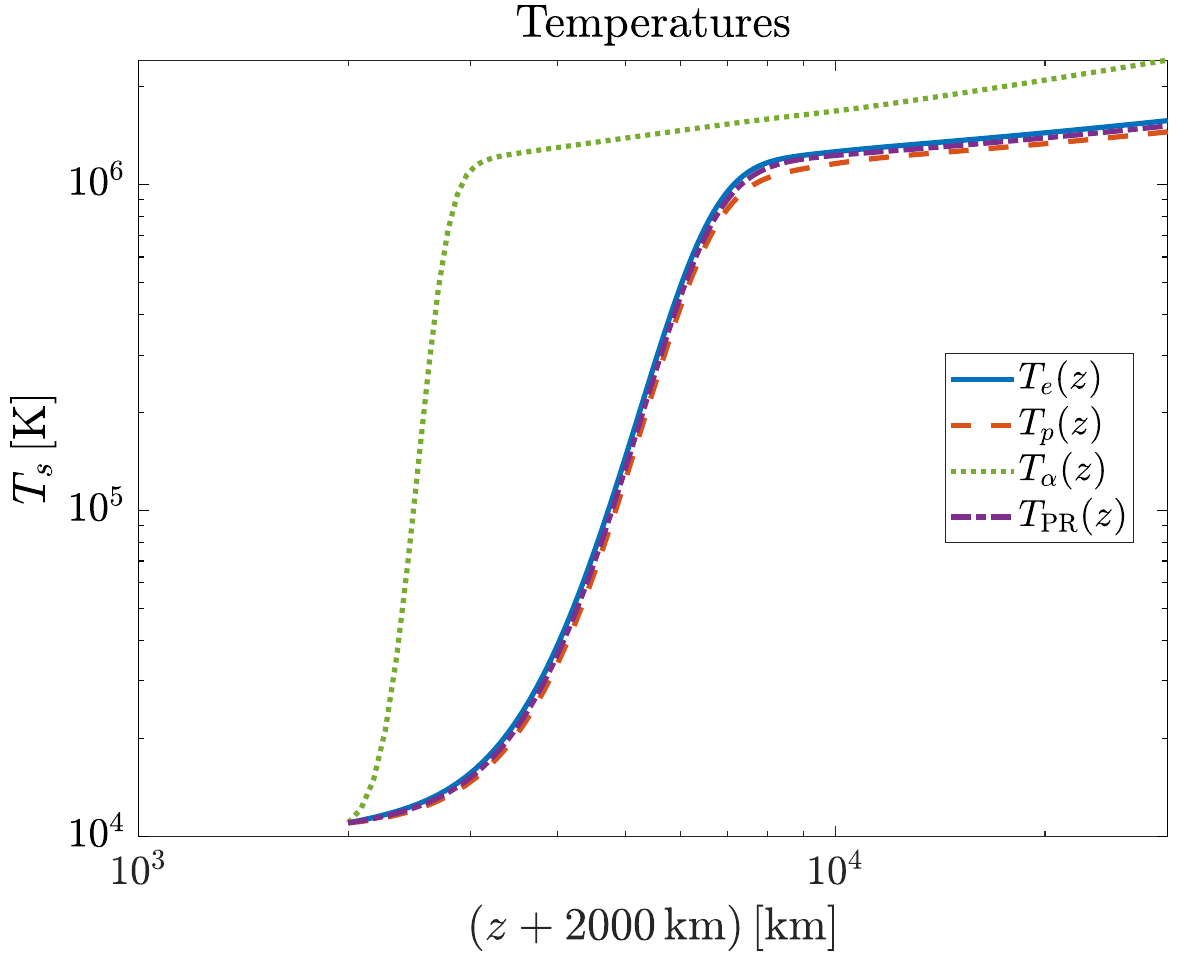}
\caption{
Effective temperatures of electrons, protons, $\alpha$ particles, and the
corresponding Pannekoek--Rosseland solution as functions of height.
The temperatures are computed from 
Equation~\eqref{paralleltemperaturepanne2} using the self-consistent density profiles shown in Figure~\ref{fig3}. The values of the physical parameters are the same as those adopted in Figure~\ref{fig3}.
}
\label{fig4}
\end{figure}

\begin{figure}
\centering
\includegraphics[width=0.5\columnwidth]{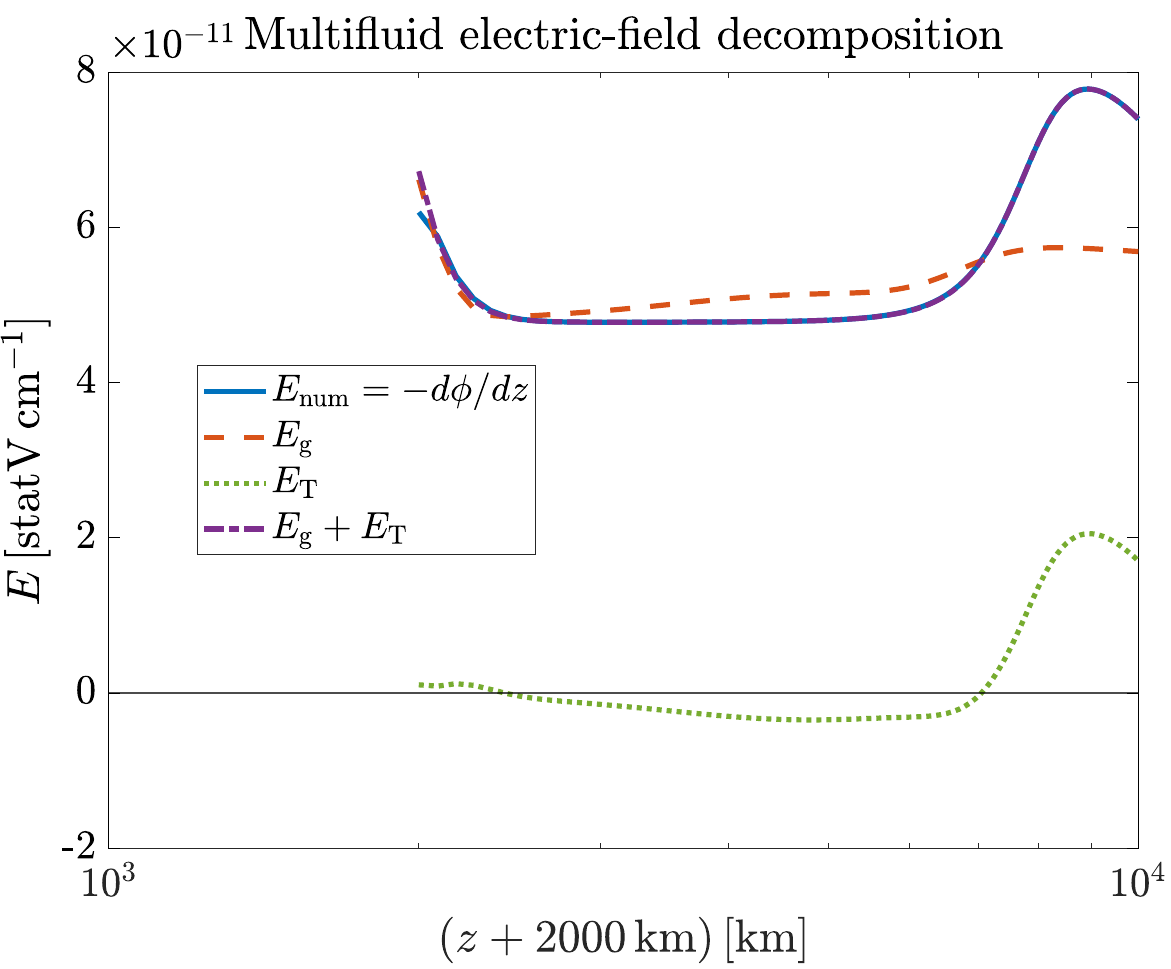}
\caption{
Decomposition of the self-consistent ambipolar electric field in the
multi-temperature atmosphere. The numerical electric field,
\(E_{\rm num}=-d\phi/dz\), is obtained by numerically solving
Equation~\eqref{eq:FM_three_species}. The gravitational contribution
\(E_{\rm g}\) is computed from
Equation~\eqref{eq:E_multi-fluid_gravitational}, the thermoelectric
contribution \(E_{\rm T}\) from
Equation~\eqref{eq:E_multi-fluid_thermal}. The values of the physical parameters are the same as those adopted in Figure~\ref{fig3}.
}
\label{fig5}
\end{figure}

\subsection{Density, Temperature, Total Potential Profiles}\label{subsec:density,temperature,total potential multi}

We now present the density, temperatures, total potentials profiles of the multi-temperature atmosphere,
highlighting the differences among the three plasma species and the differences compared to
the classical electron--proton atmosphere. Regarding the distribution of heating events,
\citet{Barbieri2025c} show that, when physical constraints are set on $\gamma(T)$, the detailed functional form of $\gamma(T)$ has only a minor influence on the resulting density and temperature profiles. Throughout this work
we therefore adopt the exponential distribution
\begin{equation}
\label{exponentialincrements}
\gamma(T)
=
\frac{1}{\Delta T}
\exp
\left[
-\frac{T-T_0}{\Delta T}
\right]
\qquad
\text{if}\qquad T>T_0.
\end{equation}
As discussed by \citet{barbieri2026}, this choice is motivated by
recent observational and numerical studies
\citep{Huang2023,Dolliou2023,Dolliou_2024,Dolliou2025}, which indicate
that the majority of heating events occur at temperatures below
$10^6\,{\rm K}$, whereas progressively hotter events become
increasingly rare. We focus on the regime where
\begin{equation}\label{physicalregime}
A_t \ll 1
\qquad \text{and} \qquad
\Delta T \gg T_0.
\end{equation}
This limit corresponds to infrequent heating events that generate temperatures far exceeding the background chromospheric temperature. In this regime, the multi-temperature model developed by \citet{Barbieri2023temperature,Barbieri2024b} successfully reproduces the observed density and temperature profiles of the solar atmosphere.

The electrostatic potential is obtained by numerically solving
Equation~\eqref{eq:FM_three_species} using a fourth-order Runge--Kutta
scheme.  Figure~\ref{fig3} summarizes the density and total potential profiles for the multi-temperature solution. The right panel shows the total potential energies experienced by the
different plasma species. As predicted by the analytical solution, the
ordering of the total potentials
satisfies Equation~\eqref{eq:potential_ordering}. The potentials are no longer
strictly linear functions of height as in the electron--proton limit and deviate from the
Pannekoek--Rosseland profile owing to the non-Maxwellian thermal
response introduced by stochastic heating. $\alpha$ particles experience
the steepest increase in the total potential energy because of their
larger mass-to-charge ratio, whereas protons remain the least
gravitationally bound species.

The left panel displays the corresponding normalized density profiles.
Their ordering is again the reverse of the potential ordering and thus consistent with Equation~\eqref{eq:ordering_densities}. As in the
single-temperature atmosphere, protons possess the largest effective
scale height, electrons occupy an intermediate position owing to the
ambipolar electric field, and $\alpha$ particles exhibit the strongest
gravitational stratification.

Unlike the single-temperature solution shown in Figure \ref{fig1}, the density profiles are no
longer simple exponential functions. Instead, they display a
characteristic two-stage decrease produced by energy-dependent
gravitational filtering. Across the lower part of the transition region, where the total
potential becomes comparable to the thermal energy of the cold
population, \(V_s(z)\sim k_B T_0\), the colder particles are
preferentially filtered out by gravity, producing the rapid initial
decrease of the density. As the colder particles become progressively depleted, the remaining
population is increasingly dominated by hotter particles, whose
gravitational scale heights are larger. Consequently, the density
gradient decreases in magnitude with height, producing a flatter
density profile above the transition region than in its lower part.

The central panel shows the corresponding density profiles. The
electron and proton densities remain close throughout the domain, as a consequence of charge neutrality and the low $\alpha$-particle abundance. In contrast, the $\alpha$-particle density exhibits a steeper decrease with height because $\alpha$-particles experience a larger total potential energy and, consequently, have a smaller gravitational scale height.
Figure~\ref{fig3} Shows that
gravitational filtering modifies the detailed shape of the density profiles without
altering their analytical ordering.

Figure~\ref{fig4} shows the corresponding temperature profiles. The temperature ordering
remains unchanged throughout the atmosphere and satisfies
Equation~\eqref{eq:temperatures_ordering}. Electrons exhibit slightly
higher temperatures than protons, whereas $\alpha$ particles
remain systematically hotter because they experience the strongest
gravitational filtering.

Unlike the single-temperature atmosphere, the effective temperatures
are no longer constant but increase monotonically with height. This
behaviour is the direct counterpart of the density profiles shown in
Figure~\ref{fig3}. As gravity progressively removes the low-energy
particles, the remaining population becomes increasingly enriched in
higher-energy particles even though no additional heating occurs with height.

The rapid temperature increase observed at relatively low altitudes is
associated with the efficient removal of the coldest particles from the
distribution. Once these particles have been filtered out, the
remaining population becomes progressively dominated by hotter
particles, and the temperature increases more gradually.
This effect is particularly pronounced for $\alpha$ particles. Their
larger mass produces a considerably steeper total potential energy than
the electrons and protons, leading to a much stronger
gravitational selection of the low-energy population. Therefore,
the temperature transition
is significantly sharper for $\alpha$ particles than for the
other plasma species.

\subsection{Multi-Fluid Electric Field Contributions}

Figure~\ref{fig5} shows the decomposition of the self-consistent
electric field into its gravitational and thermoelectric components,
defined by Equations~\eqref{eq:E_multi-fluid_gravitational} and
\eqref{eq:E_multi-fluid_thermal}.
The gravitational term dominates the ambipolar electric field in
magnitude. Near the lower boundary, the $\alpha$-particle temperature begins its
transition toward coronal values at a lower altitude than the electron
and proton temperatures, while the $\alpha$-particle density decreases
more steeply with height.
Both effects strongly reduce the weight
\(\nHe/\THe\) entering
Equation~\eqref{eq:E_multi-fluid_gravitational}. Since $\alpha$ particles carry
a relatively large mass-to-charge ratio, their preferential depletion
lowers the mass-to-charge weighting of the plasma and
produces the decrease of \(E_g\) for small $z$ values. In the limit of $\alpha$-particle contribution becomes negligible, the
gravitational term is controlled mainly by electrons and protons. Since
\(n_e\simeq n_p\), Equation \eqref{eq:E_multi-fluid_gravitational} becomes approximately
\begin{equation}
    E_g\simeq\frac{m_pg}{e}\frac{T_e}{T_e+T_p}.
\end{equation}
The nearly parallel evolution of \(T_e\) and \(T_p\) across most of the
transition region therefore accounts for the extended plateau of
\(E_g\). At larger heights, the electron temperature becomes slightly
greater than the proton temperature, producing only the modest
increase of the gravitational term for large $z$ values.

The thermoelectric term given by
Equation~\eqref{eq:E_multi-fluid_thermal} is instead determined by the
charge-weighted logarithmic temperature gradients. For electrons,
protons, and $\alpha$ particles, its numerator is proportional to
\begin{equation}
-n_e\frac{d\ln T_e}{dz}
+
n_p\frac{d\ln T_p}{dz}
+
2\,n_{\rm \He}\frac{d\ln T_{\rm \He}}{dz}.
\end{equation}
Its sign is therefore determined by the competition between the
negative electron contribution and the combined positive ion
contributions. Across the part of the transition region where
\(E_{\rm T}<0\), the density-weighted electron temperature gradient
satisfies
\begin{equation}
n_e\frac{d\ln T_e}{dz}
>
n_p\frac{d\ln T_p}{dz}
+
2\,n_{\rm \He}\frac{d\ln T_{\rm \He}}{dz},
\end{equation}
so that the negative electron contribution dominates. On the upper side of the transition region, the electron temperature
begins to level off while the proton temperature continues to
increase. Since each contribution to \(E_{\rm T}\) is weighted by the
corresponding species density, the change in sign cannot be attributed
to the temperature gradients alone. Rather, the combined evolution of
the densities and temperature gradients eventually leads to
\begin{equation}
n_p\frac{d\ln T_p}{dz}
+
2\,n_{\rm \He}\frac{d\ln T_{\rm \He}}{dz}
>
n_e\frac{d\ln T_e}{dz}.
\end{equation}
The positive ion contribution then exceeds the negative electron
contribution, causing \(E_{\rm T}\) to reverse sign and increase
rapidly. The pronounced maximum of the total electric field is
therefore primarily thermoelectric in origin. At still greater
heights, the temperature profiles progressively flatten, reducing the
density-weighted logarithmic temperature gradients and hence
\(E_{\rm T}\). This accounts for the slight decline of the total
electric field beyond its maximum.

The non-monotonic behaviour of the ambipolar electric field therefore
does not originate from the gravitational contribution alone, but from
its coupling with the thermoelectric correction generated by the
species-dependent temperature gradients. The multi-temperature
atmosphere thus differs from the classical Pannekoek--Rosseland
equilibrium not only because the gravitational contribution is
modified, but also because gravitational filtering generates an
additional thermoelectric contribution that reshapes the ambipolar electric field.

\subsection{Perturbative Solution for a Small $\alpha$-particle Abundance}
\label{sec:small_fb_expansion}

As shown in Figure~\ref{fig3}, the multi-temperature solution does not satisfy the small-potential condition given by Equation~\eqref{eq:small_potential_condition}. As a consequence, the perturbative expansion developed in Subsection \ref{subsec:local_analytical_solution} is no longer applicable. Nevertheless, an analytical approximation can still be obtained directly from the charge-neutrality condition by exploiting the small $\alpha$-particle abundance observed in the solar corona. Since the abundance ratio
\begin{equation}
f_b=\frac{n_{0,\rm \He}}{n_{0,p}}
\ll1,
\end{equation}
is typically only a few percent, the electrostatic potential can be expanded perturbatively as
\begin{equation}
\phi(z)
\simeq
\phi_0(z)
+
f_b\,\phi_1(z),
\label{eq:phi_fb_expansion_def}
\end{equation}
while retaining the full dependence of the multi-temperature distribution. We introduce the function
\begin{equation}
\begin{aligned}
\mathcal{Q}(V)
={}&
A_t\int_{T_0}^{+\infty}
\gamma(T)
\exp\left(-\frac{V}{k_BT}\right)dT+
(1-A_t)
\exp\left(-\frac{V}{k_BT_0}\right),
\end{aligned}
\label{eq:def_Q}
\end{equation}
so that the density of each species can be written in the form
\begin{equation}
n_\alpha(z)
=
n_{0,\alpha}
\mathcal{Q}\!\left[V_\alpha(z)\right].
\end{equation}

Using the boundary condition
\begin{equation}
n_{0,e}=n_{0,p}+2n_{0,\rm \He},
\end{equation}
the charge-neutrality condition, Equation~\eqref{eq:neutrality}, becomes
\begin{equation}
(1+2f_b)\mathcal{Q}(V_e)
=
\mathcal{Q}(V_p)
+
2f_b\mathcal{Q}(\VHe).
\label{eq:QN_Q}
\end{equation}

At zeroth order in \(f_b\), Equation~\eqref{eq:QN_Q} reduces to
\begin{equation}
\mathcal{Q}\!\left(V_e^{(0)}\right)
=
\mathcal{Q}\!\left(V_p^{(0)}\right).
\end{equation}
Since the function \(\mathcal{Q}(V)\) is monotonically decreasing with increasing total potential $V$, this equality immediately demands that
\begin{equation}
V_e^{(0)}
=
V_p^{(0)}
\equiv
V_0.
\end{equation}
Thus, to leading order, electrons and protons experience the same total potential as in the classical Pannekoek--Rosseland equilibrium. The corresponding electrostatic potential is therefore
\begin{equation}
\phi_0(z)
=
-\frac{m_p-m_e}{2\,e}g\,z,
\label{eq:phi0_fb}
\end{equation}
which is identical with the classical Pannekoek--Rosseland solution given in Equation~\eqref{eq:PRpotential}. The common electron--proton total potential is then
\begin{equation}
V_0(z)
=
\frac{m_p+m_e}{2}g\,z,
\label{eq:V0_fb}
\end{equation}
whereas the $\alpha$ particles experience
\begin{equation}
\VHe^{(0)}(z)
=
(3\,m_p+m_e)\,g\,z.
\label{eq:VHe0_fb}
\end{equation}
These zeroth-order potentials already provide a direct physical
interpretation of the species-dependent profiles shown in
Figures~\ref{fig3} and~\ref{fig4}. For a population at a given
temperature \(T\), the characteristic vertical scale associated with
gravitational filtering is inversely proportional to the total
potential gradient,
\begin{equation}
H_s(T)
\sim
\frac{k_B T}{dV_s^{(0)}/dz}.
\end{equation}
Electrons and protons have the same total potential at zeroth order and
therefore the same characteristic scale for any \(T\). By contrast,
neglecting \(m_e\),
\begin{equation}
\frac{dV_{\rm \He}^{(0)}/dz}{dV_0/dz}
\simeq 6,
\end{equation}
so that the corresponding $\alpha$-particle scale is approximately six
times smaller,
\begin{equation}
H_{\rm \He}(T)
\simeq
\frac{1}{6}H_{e,p}(T).
\end{equation}
This relation holds for every component of the multi-temperature
distribution, including the background population at \(T=T_0\).
Consequently, gravitational filtering acts over a substantially
shorter vertical interval for $\alpha$-particles, producing the sharper density and temperature transitions seen in
Figures~\ref{fig3} and~\ref{fig4}. Conversely, the equality of the electron and proton zeroth-order potentials explains why their density and temperature profiles remain close to each other. Finite $\alpha$-particle abundance introduces differences between the electron and proton profiles only through higher-order corrections in \(f_b\). These zeroth-order relations therefore provide an analytical basis for the qualitative interpretation of the density, temperature, and total
potential profiles discussed in Section~\ref{subsec:density,temperature,total potential multi}.
The leading-order ambipolar field is independent of the stochastic-heating parameters and remains identical to the classical Pannekoek--Rosseland field. The effects of the multi-temperature distribution only appear through the first-order correction proportional to \(f_b\). To determine this correction, the electron and proton total potentials are expanded as
\begin{equation}
V_e
\simeq
V_0-f_b\,e\,\phi_1
\end{equation}
\begin{equation}
V_p
\simeq
V_0+f_b\,e\,\phi_1.
\end{equation}
The first-order correction to the $\alpha$-particle potential does not contribute, since its contribution to Equation~\eqref{eq:QN_Q} is already proportional to \(f_b\). Expanding Equation~\eqref{eq:QN_Q} to first order in \(f_b\) therefore yields
\begin{equation}\label{eq:solvingQ}
2\mathcal{Q}(V_0)
-
2\,e\,\phi_1 \mathcal{Q}'(V_0)
-
2\mathcal{Q}\!\left(\VHe^{(0)}\right)
=
0,
\end{equation}
where $\mathcal{Q}'(V) = \partial Q / \partial V $. Solving for the first-order correction gives
\begin{equation}
\phi_1(z)
=
\frac{
\mathcal{Q}\!\left[V_0(z)\right]
-
\mathcal{Q}\!\left[\VHe^{(0)}(z)\right]
}{
e\,\mathcal{Q}'\!\left[V_0(z)\right]
},
\label{eq:phi1_fb}
\end{equation}
and the electrostatic potential becomes
\begin{equation}
\phi(z)
\simeq
-\frac{m_p-m_e}{2\,e}g\,z
+
\frac{f_b}{e}
\frac{
\mathcal{Q}\!\left[\dfrac{m_p+m_e}{2}g\,z\right]
-
\mathcal{Q}\!\left[(3\,m_p+m_e)g\,z\right]
}{
\mathcal{Q}'\!\left[\dfrac{m_p+m_e}{2}g\,z\right]
}.
\label{eq:phi_multi-temperature_fb}
\end{equation}

Equation~\eqref{eq:phi_multi-temperature_fb} constitutes an analytical
approximation valid to first order in the $\alpha$-particle abundance while
retaining the full multi-temperature dependence of the particle
distribution. Unlike the solution derived in Section \ref{subsec:local_analytical_solution},
no expansion of the Boltzmann factors with respect to the
total potential has been introduced. Consequently, the approximation
remains valid even when the total potentials become comparable to or
greater than the characteristic thermal energy, provided that the
$\alpha$-particle abundance satisfies \(f_b\ll1\).

The first-order correction is entirely determined by the functions \(\mathcal{Q}\) of the electron--proton plasma and the alpha
particles. Since
\(
\VHe^{(0)}>V_0,
\)
$\alpha$ particles experience a stronger gravitational filtering than
electrons and protons. The electrostatic potential must therefore
increase with respect to the classical Pannekoek--Rosseland solution in
order to preserve charge-neutrality throughout the atmosphere. As a consistency check, the small-potential limit of
Equation~\eqref{eq:phi1_fb} can be recovered by expanding the thermal
response around \(V=0\):
\begin{equation}
\mathcal{Q}(V)
\simeq
\mathcal{Q}(0)
+
V\mathcal{Q}'(0).
\end{equation}
Substituting this approximation into Equation~\eqref{eq:phi1_fb} yields
\begin{equation}
e\,\phi_1
\simeq
V_0-\VHe^{(0)}
=
-\frac{5m_p+m_e}{2}g\,z.
\end{equation}
Neglecting the electron mass, the electrostatic potential reduces to
\begin{equation}
\phi(z)
\simeq
-\frac{m_p\,g}{2\,e}
\left(1+5f_b\right)z.
\label{eq:phi_small_fb_small_V}
\end{equation}
This expression corresponds to the first-order expansion of the
single-temperature solution, Equation~\eqref{eq:local_phi_effective_mass},
in the limit \(m_e\ll m_p\) and
\(\mHe \simeq4\,m_p\) since:
\begin{equation}
\frac{1+8f_b}{1+3f_b}
\simeq
1+5f_b.
\end{equation}
The agreement between these two
approaches provides a consistency check of the analytical
multispecies formulation and shows that  the two expansions are
complementary. The small-potential approximation is appropriate close
to the base of the atmosphere, where the total potentials remain much
smaller than the characteristic thermal energy, whereas the present
expansion, defined by Equation \eqref{eq:phi_fb_expansion_def}, remains applicable throughout the atmosphere without
requiring \(V_\alpha/k_BT\ll1\), provided that the $\alpha$-particle abundance
remains small.

\begin{figure}
\centering
\includegraphics[width=0.5\columnwidth]{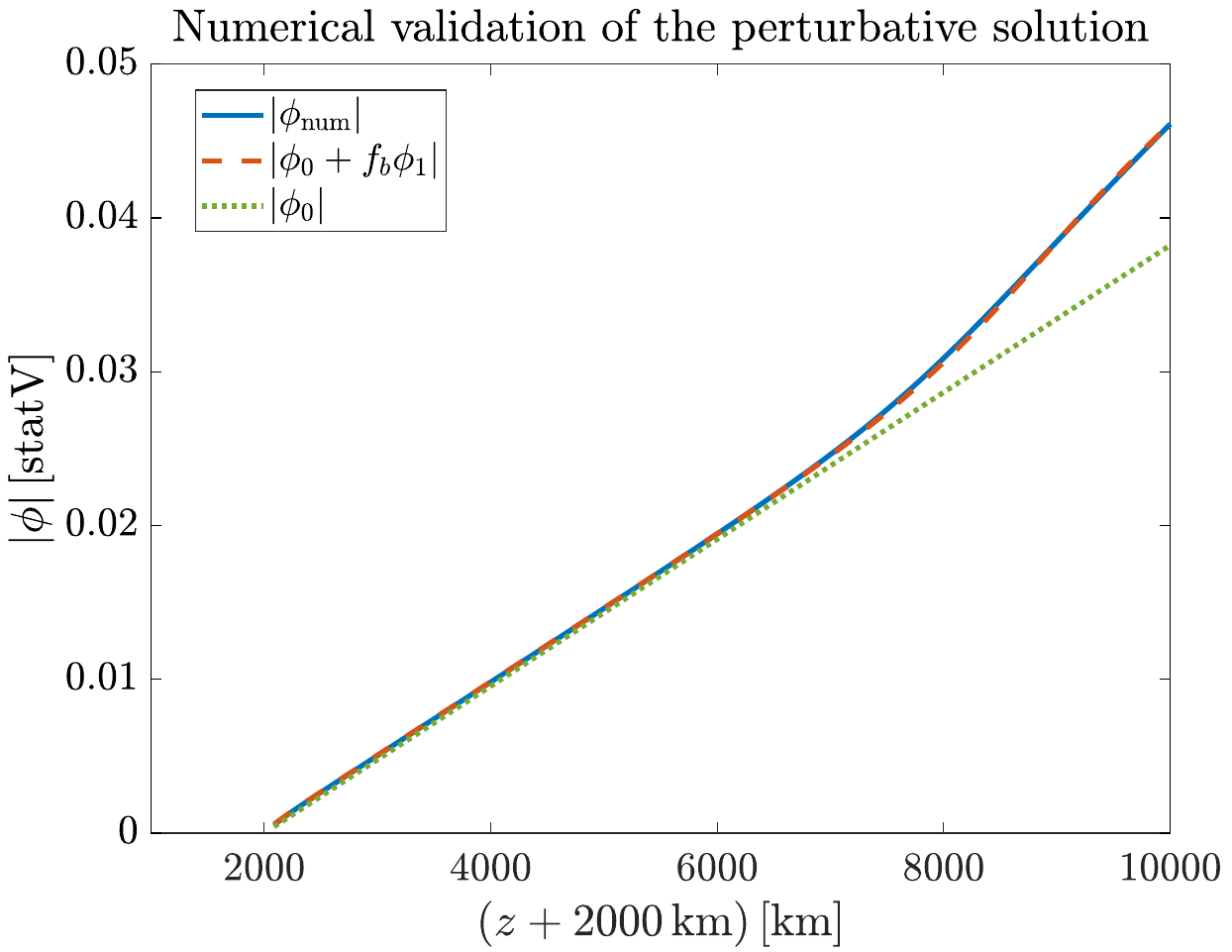}
\caption{
Comparison between the numerical electrostatic potential obtained by
solving Equation~\eqref{eq:FM_three_species}, the analytical first-order
perturbative solution,
\(\phi\simeq\phi_0+f_b\phi_1\), given by
Equation~\eqref{eq:phi_multi-temperature_fb} and the leading order perturbative solution given by Equation \eqref{eq:phi0_fb}. The numerical values of the
parameters are the same as in Figure~\ref{fig3}.
}
\label{fig6}
\end{figure}

To assess the accuracy of the perturbative expansion, the analytical
solution given by Equation~\eqref{eq:phi_multi-temperature_fb} is compared
with the numerical solution of Equation~\eqref{eq:FM_three_species} in
Figure~\ref{fig6}. Also shown is the zeroth-order solution,
\(\phi_0\). While \(\phi_0\) provides a good
approximation in the lower atmosphere, it progressively deviates from
the numerical solution at larger heights, particularly near the top of
the transition region where stochastic heating becomes increasingly
important. In contrast, the first-order perturbative solution,
\(\phi\simeq\phi_0+f_b\phi_1\), remains almost indistinguishable from
the numerical result over the entire computational domain. This
excellent agreement demonstrates that the first-order correction
captures the leading effects of stochastic heating while fully
retaining the multi-temperature nature of the plasma.

\subsection{Species-Dependent Width of the Transition Region}

The perturbative solution derived above provides a simple estimate
of the characteristic width of the transition region for each plasma
species. Following the definition introduced by
\citet{Barbieri2025c}, the lower boundary of the transition region is
defined as the height at which the hot and cold contributions to the
numerator of the effective temperature in
Equation~\eqref{paralleltemperaturepanne2} become equal, while its upper
boundary is defined analogously from the denominator.

As discussed in Sec. \ref{sec:small_fb_expansion}, the electrostatic potential
is well approximated by the perturbative solution
\(\phi\simeq\phi_0+f_b\phi_1\). Since the first-order correction
remains small, the total potentials vary almost linearly across the
transition region (Figure~\ref{fig3}), so that each species can be
locally approximated as

\begin{equation}
V_s(z)
\simeq
K_s z,
\end{equation}
where
\begin{equation}
K_s
=
\left.
\frac{dV_s}{dz}
\right|_{\rm TR}
=
m_sg
-
e_sE_{\rm TR},
\end{equation}
and \(E_{\rm TR}\) denotes the ambipolar electric field evaluated
across the transition region. Proceeding as \citet{Barbieri2025c}, the transition-region width
becomes
\begin{equation}
\Delta z_s
\simeq
\frac{k_BT_0}{K_s}
\ln\left(
\frac{\langle T\rangle_\gamma}{T_0}
\right),
\label{eq:TR_width_multispecies}
\end{equation}
where
\begin{equation}
\langle T\rangle_\gamma
=
\int_{T_0}^{+\infty}
T\gamma(T)\,dT
\end{equation}
is the mean temperature of the stochastic-heating events. Equation~\eqref{eq:TR_width_multispecies} shows that the width of the
transition region is inversely proportional to the gradient of the
total potential experienced by each plasma species. Since the
perturbative electrostatic potential remains close to the
Pannekoek--Rosseland solution, the ordering of the total potential
gradients is preserved:
\begin{equation}
K_p
<
K_e
<
K_{\rm \alpha},
\end{equation}
which implies that
\begin{equation}
\Delta z_{\rm \alpha}
<
\Delta z_e
<
\Delta z_p.
\end{equation}

$\alpha$ particles therefore undergo the transition from the cold
chromospheric population to the suprathermal coronal population over
the shortest vertical distance of all particle species under investigation. Their larger mass-to-charge ratio
produces the steepest total potential, leading to the most efficient
gravitational filtering of low-energy particles. Conversely, protons
experience the shallowest total potential and therefore exhibit the
widest transition region, while electrons occupy an intermediate
position close to protons. This analytical prediction is fully consistent with the
density and temperature profiles shown in
Figure~\ref{fig3}, where the transition associated with $\alpha$ particles
is significantly steeper than those of electrons and protons.

\section{Summary and Perspectives}  \label{sec:conclusions}

In this work, we develop a kinetic model of gravitationally stratified multispecies plasma atmospheres composed of electrons, protons, and $\alpha$ particles, 
extending the classical Pannekoek--Rosseland equilibrium to both multi-species and multi-temperature plasmas.

For the single-temperature atmosphere, we derive a general self-consistent expression for the ambipolar electric field and show that its strength is entirely controlled by the local plasma composition. In particular, the inclusion of $\alpha$ particles enhances the ambipolar electric field because of their larger mass-to-charge ratio, leading to an analytical ordering of the total potential energies and density profiles of the different species. We also obtain zeroth-order and
first-order analytical solutions for the electrostatic potential, demonstrating excellent agreement between the analytical approximation and the full numerical solution.

We then generalize the kinetic model to a multi-temperature plasma produced by stochastic boundary heating. The resulting stationary
distribution is a superposition of Maxwellian populations and gives rise to self-consistent density and effective temperature profiles that depend on the statistical properties of the heating process. Although
the detailed shape of the density and temperature profiles is modified by stochastic heating, the analytical ordering of the total potentials, densities, and temperatures  of the different species remains unchanged compared to the single-temperature case.

Stochastic heating produces gravitational filtering of the particle populations. Since low-energy particles are preferentially removed with increasing altitude, the remaining plasma becomes progressively enriched in
higher-energy particles. As a consequence, the density profiles depart from simple exponential stratification and the temperatures increase monotonically with height despite the absence of any local heating process.

The multi-fluid analytical formulation further provides a transparent physical interpretation of the kinetic solution. The ambipolar electric field can be decomposed into a gravitational contribution, already present in the classical Pannekoek--Rosseland equilibrium, and an additional thermoelectric contribution generated by species-dependent temperature gradients. While the gravitational component remains the dominant contribution throughout the atmosphere, the thermoelectric field is responsible for the non-monotonic behaviour of the total ambipolar electric field through the temperature gradients generated by
gravitational filtering.

Since the local small-potential expansion is no longer valid
in the multi-temperature case, we derive a complementary analytical
solution based on the assumption of a small  $\alpha$-particle abundance (\(f_b\ll1\)). This
perturbative approach preserves the full multi-temperature dependence of
the plasma and accurately reproduces the numerical electrostatic
potential, thereby extending the analytical description of ambipolar
equilibria beyond the regime where the small-potential approximation
applies.

Finally, the multispecies formulation also provides a simple analytical
estimate of the transition-region width associated with each plasma
species. Owing to the different total potentials experienced by
electrons, protons, and $\alpha$ particles, the model predicts
species-dependent transition-region widths, with $\alpha$ particles
undergoing the transition over the shortest vertical distance.

The strong alpha-particle stratification obtained in the present model
should be interpreted in the context of the collisionless approximation
adopted here. Observations indicate that the helium abundance relative
to hydrogen can vary substantially in the solar corona and solar wind,
with significant helium depletion occurring in some coronal structures
\citep[e.g.,][]{Moses2020}. The preferential depletion of alpha
particles predicted by our model is therefore qualitatively consistent
with the observed tendency for helium to become depleted relative to
hydrogen. However, the comparatively narrow alpha-particle transition
obtained in Fig.~\ref{fig3} is not, to our knowledge, a directly established
observational property of the solar transition region. In particular,
current observations do not directly constrain the relative transition
widths of alpha particles, protons, and electrons in the form predicted
by our model. Its pronounced species-dependent stratification
is therefore a prediction of the
collisionless limit rather than as a quantitatively established
property of the solar atmosphere.

The present model neglects Coulomb collisions, which are known to play
an important role in shaping the density and temperature profiles of
multitemperature plasmas \citep[e.g.,][]{barbieri2026}. In a multispecies plasma,
collisional momentum and energy exchange couple the evolution of the
different species and may therefore modify the differential
stratification obtained in the collisionless limit. In particular,
multifluid studies have shown that sufficiently strong collisional
coupling can drive different species toward a more collective,
single-fluid-like stratification, whereas significant species
decoupling persists when the coupling becomes weaker
\citep[e.g.,][]{Zhang2026}. Coulomb collisions can therefore reduce
the pronounced difference between the alpha-particle and proton scale
heights, although the magnitude of this effect cannot be
determined with our collisionless model.

Our work is a first step toward a
self-consistent kinetic description of multicomponent plasma
atmospheres subject to stochastic heating. Given the nature of the
present results, we expect Coulomb collisions to affect all plasma
species, modifying both their density and effective temperature
profiles. An important extension of the present work will therefore be
the generalization of the collisional formalism developed by
\citet{barbieri2026} to the multispecies atmosphere considered here.
Such a formulation will enable investigations of the impact of
collisions on the stratification of the different plasma species and
their contributions to both the gravitational and thermoelectric
components of the ambipolar electric field.

\begin{acknowledgments}
The authors wish to thank the anonymous reviewer for the valuable comments and suggestions that helped improve the clarity and quality of this work. L.B. wants to thank Sorbonne Universit\'e for financial support within the framework of the Initiative Physique des Infinis. D.V. is supported by STFC Consolidated Grant ST/W/001004/1.
\end{acknowledgments}

\section*{Conflict of interest}
The authors have no conflicts to disclose.

\bibliography{biblio}{}
\bibliographystyle{aasjournal}

\end{document}